\documentclass[12pt]{article}
\usepackage{amsmath,amssymb,amsfonts,amsthm}
\title{Superpositions of the dual family of nonlinear coherent states and their non-classical properties}
\author{O Abbasi and M K Tavassoly
\\
\footnotesize{Atomic and Molecular Group, Faculty  of Physics,
Yazd University, Yazd, Iran}
\\ \footnotesize{e-mail: mktavassoly@yazduni.ac.ir  } }

\begin{document}


\newcommand{\I}{\mathbb{I}}
\newcommand{\norm}[1]{\left\Vert#1\right\Vert}
\newcommand{\abs}[1]{\left\vert#1\right\vert}
\newcommand{\set}[1]{\left\{#1\right\}}
\newcommand{\R}{\mathbb R}
\newcommand{\C}{\mathbb C}
\newcommand{\DD}{\mathbb D}
\newcommand{\eps}{\varepsilon}
\newcommand{\To}{\longrightarrow}
\newcommand{\BX}{\mathbf{B}(X)}
\newcommand{\HH}{\mathfrak{H}}
\newcommand{\D}{\mathcal{D}}
\newcommand{\N}{\mathcal{N}}
\newcommand{\W}{\mathcal{W}}
\newcommand{\RR}{\mathcal{R}}
\newcommand{\HD}{\hat{\mathcal{H}}}

 \maketitle


 \begin{abstract}
  Nonlinear coherent states (CSs) and their {\it dual families} were introduced recently.
  In this paper  we want to obtain their superposition and investigate their non-classical
  properties such as antibunching effect, quadrature squeezing
  and amplitude squared squeezing.
  For this purpose two types of superposition are considered. In the first type we neglect the
  normalization factors of the two components of the dual pair, superpose them and then we normalize
  the obtained states, while in the second type we superpose the two normalized components and then
  again normalize the resultant states.
  As a physical realization, the formalism will then be applied to a special physical system with
  known nonlinearity function, i.e., Hydrogen-like spectrum. We continue with the (first type of)
  superposition  of the dual pair of Gazeau-Klauder coherent states (GKCSs) as  temporally stable CSs.
  An application of the proposal will be given by employing the P\"{o}schl-Teller potential system.
  The numerical results are presented and discussed in detail,  showing the effects of this special
  quantum interference.
 \end{abstract}


 {\bf keywords:}
    nonlinear coherent states;  dual family of nonlinear coherent states;  superposition of states.


  \section{Introduction}\label{sec-intro}
  Following the development of quantum theory of radiation field, the notion of
  {\it standard coherent states} (CSs) as the states that are most nearly close
  to classical description of radiation field, were introduced \cite{glauber1, Sudarshan}.
  These states contain important concepts in quantum optics and have numerous applications
  in different branches of modern physics \cite{klauder-book,ali book,application}.
  The standard CSs of a radiation field are usually constructed using one of the three following manners:
  i) they are eigenstates of the standard annihilation operator of the harmonic oscillator, i.e.,
  $a \vert\alpha\rangle=\alpha\vert\alpha\rangle$,
  ii) they can be reproduced by the action of a unitary displacement operator on the vacuum of radiation field, i.e.,
  $\exp (\alpha a^{\dag}-\alpha^{\ast} a)\vert0\rangle=\vert\alpha\rangle$.
  iii) they minimize the Heisenberg uncertainty relation ${(\Delta x)^2}{(\Delta p)^2}\;\geq\frac{1}{16}$,
  with equal variances as the limit of vacuum in both quadratures.
  In recent years CSs have been generalized in various ways.
  One of the most important
  generalization based on algebraic aspects is the notion of {\it nonlinear CSs} \cite{Matos1996} or
  {\it $f$-CSs} \cite{Manko1997}.
  These states defined as the right eigenstates of an $f$-deformed annihilation operator $A=af(n)$,
  where $f$ is an operator-valued function of number operator $n=a^{\dag}a$ characterizes the nonlinearity
  nature of the system.
  The nonlinear CSs approach provides a powerful method for classification  and unification of a vast classes
  of generalized CSs \cite{Roknizadeh2004}. They are also useful in the construction of CSs associated to
  inverse bosonic and $f$-deformed annihilation operators \cite{mktinverse}.
  A question may naturally arise: is there any displacement operator whose the action on the vacuum state
  can reproduce nonlinear CSs? It is shown that the answer is affirmative. It is possible to define a
  {\it displacement type} operator using the auxiliary operators $B=a\;\frac{1}{f(n)}$ and its conjugate
  $B^{\dag}$. Indeed it is proved that two different displacement type  operators may be constructed.
  While the action of one of them on the vacuum reproduce the original nonlinear CSs, the other
  one produces a new set of nonlinear CSs \cite{roy roy} known as the {\it dual pair} with respect
  to each other \cite{ART}.

  On the other side in recent years much attention has been paid to the {\it quantum
  superposition}
  of CSs \cite{ yurk, advance} and nonlinear CSs \cite{superpos new} due to their
  interference effects \cite{domokos}. Even and odd CSs \cite{dodonov, Schleish} (even and odd nonlinear
  CSs \cite{mancini}) respectively as symmetric and anti-symmetric superposition of CSs
  (nonlinear CSs) are along this subject of researches. Generally, interference effects in f-deformed fields were discussed in \cite{mancini2}.
  The generation of non-classical states possessing properties such as sub-Poissonian behavior,
  quadrature squeezing and higher order squeezing is the main motivation of this subject of researches.

  As the main goal of the present paper we want to construct the general formalism of superposition of the dual pair
  corresponding to any nonlinear CS class with known $f(n)$.
  For this purpose two types of superposition are considered. In the first type we
  neglect the normalization factors of the dual family of nonlinear CSs, superpose
  them and then normalize the obtained states. To distinguish this kind of superposition
  from the second one we call it as "{\it combination}". It will be easily shown that this
  combination is a kind of "superposition of the normalized states" with different weights,
  where the weights relate to the normalization factors of the individual states (see equation (11)
  of the paper). But in the second type we superimpose the two "normalized components" and then again
  normalize the resultant states.
  It may be recognized that the second type is more customary in the literature.
  Altogether, some words seem to be necessary about the introduction of the first type.
  This combination kind of superposition has been done previously in the literature for different
  purposes \cite{prakash1, prakash2, prakash3, Wang, Nguyen}. The most general case of superposition of
  two canonical CSs $|\alpha \rangle$,  $|\beta \rangle$  is introduced as
  $|\psi\rangle = c_1|\alpha \rangle + c_2|\beta \rangle$, where $c_1$ and $c_2$ are
  complex numbers restricted by the normalization constraint.
  Squeezing of these states is investigated by considering $\langle \psi|(\Delta X_\theta)^2|\psi\rangle$,
  where $X_\theta =X_1 \cos \theta + X_2 \sin \theta$, and $X_1 +i X_2=a$
  is the annihilation operator \cite{prakash1}.
  Searching for maximum simultaneous squeezing and antibunching effects of these states
  has been done in \cite{prakash2}.
  Later, the latter states have been again studied by the same authors in \cite{prakash3} for
  investigating the (maximum) amplitude squared squeezing of the hermition
  operator $Z_\theta =Z_1 \cos \theta +Z_2 \sin \theta$, where $Z_{1,2}$ operators
  are defined by $Z_1+i Z_2 = [a - \langle \psi  |a| \psi \rangle]^2$.
  A rather different motivation for this kind of superposition may outcome from the teleportation scheme.
  Teleportation  of two-mode CSs  studied  in \cite{Wang}. The authors  considered
  a particular form of the two-mode CSs of the form $ |\psi(x, y) \rangle =
  x | \alpha , \alpha \rangle + y | -\alpha, -\alpha \rangle$, where $x, y$ are
  coefficients constrained by the normalization condition. Interestingly, such a
  particular two-mode CSs can always be obtained by first the teleportation a
  single-mode CSs of the form $|\psi(x, y) \rangle =
  x | \alpha \sqrt 2 \rangle + y | -\alpha \sqrt 2\rangle $ and
  then superposing it on a $50:50$ beam-splitter with the vacuum
  state $|0\rangle$ (see \cite{Nguyen} and references therein).
  According to our terminology, in all of the latter mentioned
  cases the "combinations of states" have been considered and studied.
  In addition to these,  the first type has this mathematical property
  that can exactly be classified in the  "nonlinear CSs"
  family with a particular nonlinearity function.

  In what follows a brief review on the $f$-CSs and their dual pairs required
  for our further manipulations is presented in section 2.
  Some explanations about the resolution of the
  identity and the domain of the dual family of states are presented in section 3.
  Then in section 4 the combination of the dual pair are constructed and the
  nonlinearity signature of the combination of states is established through
  driving the corresponding nonlinearity function.
  In section 5 the (second kind of) superposition of the dual pair has been
  introduced. The non-classical criteria needed for our further discussion
  will be introduced in section 6. Section 7 deals with a particular physical
  realization of the presented formalism by applying to  "Hydrogen-like spectrum".
  In section 8 we deal with the Gazeau-Klauder CSs (GKCSs) as the temporally stable
  CSs \cite{GK}. The combination of these states with their
  dual family will be introduced and their non-classical properties are studied and
  the nonlinearity signature of the combination of GKCSs is established.
  A summary of the results is presented in section 9. Finally, in the appendix
  some expectation values were required for our numerical calculations can be found.


 \section{The dual family of nonlinear CSs}

  The nonlinear CSs method is based on the deformation of standard annihilation and creation operators
  with an intensity dependent function $f(n)$, according to the relations \cite{Matos1996, Manko1997}
 \begin{equation}\label{A,Adag }
  A=af(n),\qquad A^{\dag}=f^{\dag}(n)a^{\dag},
 \end{equation}
 where $a$, $a^\dag$ and $n=a^{\dag} a$ are bosonic annihilation, creation and number operators, respectively.
 The commutators between $A$ and $A^{\dag}$ read as:
 \begin{equation}\label{commutators}
   [A,A^{\dag}]=(n+1)f^{\dag}(n+1)f(n+1)-nf^{\dag}(n)f(n).
 \end{equation}
  Now we choose $f(n)$ to be real,  i.e., ${f^{\dag}}(n)=f(n)$
   and nonnegative until we deal with GKCSs in section 5.
   Nonlinear CSs  satisfy the typical eigenvalue equation
 \begin{equation}\label{eigenvalue eq}
   A|\alpha,f\rangle=\alpha|\alpha,f\rangle,\qquad  \alpha\in \mathbb{C}
 \end{equation}
 The Fock space representation of these states is explicitly expressed as
  \begin{equation}\label{NCSs1}
    |\alpha,f\rangle=N_{f}\sum_{n=0}^\infty
    \frac {\alpha^n}{\sqrt{n!}\;[f(n)]!}\;| n  \rangle,
  \end{equation}
  where $N_{f}$ is some appropriate normalization factor determined from $\langle \alpha,f|\alpha,f\rangle=1$.
  By convention
 $[f(n)]!\;\doteq\;f(1)f(2)...f(n)$ and       $[f(0)]!\;\doteq\;1.$
   On the other hand by considering (\ref{commutators}) we see that $[A,A^{\dag}]$
   is neither $c$-number nor linear in the generator $n$ and therefore based on
   BCH theorem \cite{BCH} a displacement operator does not exist as
   $\exp(\alpha A^{\dag}-\alpha^{\ast} A)$ to construct nonlinear CSs.
   But in \cite{roy roy, ART} the operator $B$ and it's conjugate $B^{\dag}$ were defined as follow
 \begin{equation}\label{B Bdag}
   B=a\;\frac{1}{f(n)},\qquad      B^{\dag}=\frac{1}{f(n)}\;a^{\dag},
 \end{equation}
  which satisfy the canonical commutation relations $[A,B^{\dag}]=\hat{I}=[B,A^{\dag}]$.
  Now we can consider two displacement-type operator $D(\beta)=\exp(\beta B^{\dag}-\beta^{\ast}A)$ and
   $\tilde D(\beta)=\exp(\beta A^{\dag}-\beta^{\ast}B)$ and apply them on the vacuum state of radiation field.
   The results are respectively the states in (\ref{NCSs1}) and a new class of nonlinear CSs which named
   the dual pair of (\ref{NCSs1}) with the following expansion \cite{roy roy, ART}
 \begin{equation}\label{dual}
   |\widetilde{\beta,f}\rangle =
    \tilde {D}(\beta)|0\rangle=\tilde{N}_f\sum_{n=0}^\infty\frac {\beta^n\;
    [f(n)]!}{\sqrt{n!}}\;| n  \rangle,\qquad         \beta\in \mathbb{C},
 \end{equation}
   where again the normalization factor $\tilde{N}_f$ can be determined from
    $\langle \widetilde{\beta,f}|\widetilde{\beta,f}\rangle=1$. It
    is clear that $|\widetilde{\alpha, f}\rangle =
    |\alpha, f^{-1}\rangle$. Therefore, $|\alpha, f\rangle$ and
    $|\widetilde{\alpha, f}\rangle$ as a dual pair may be generated from two
    different types of interaction of atom-field,
    due to the fact that each of them corresponds to a distinct nonlinearity function
    which mathematically are inverse of each other.
    The dual pair which will constitute the two components of our further
    superposition in the present paper
    are not orthogonal, with the overlap
 \begin{equation}\label{overlap}
   {\langle \alpha,f} |\widetilde{\alpha,f}\rangle = N_f \tilde N_f \exp (|\alpha|^2).
 \end{equation}

 \section{Resolution of the identity for the dual family of nonlinear CSs}

 Requiring the over-completeness of any class of CSs, the dual pair of CSs should satisfy
 the resolution of the identity, in addition to the nonorthogonality condition.
 But the nonorthogonality of the dual family of states  (i.e.,  $\langle \alpha, f|\alpha', f\rangle \neq 0$ and similarly for the dual pair)
 is so obvious subject that we pay no attention to it.
 Moreover, the most important requirement for these states is the necessity  to satisfy  the resolution of the identity, i.e.,
   \begin{equation}\label{res-id}
      \int _D d^2 \alpha |\alpha, f\rangle W(|\alpha|^2)\langle \alpha, f | = \sum_{n=0}^\infty |n\rangle  \langle n|
      =\hat I
   \end{equation}
   where $d^2 \alpha \doteq dx dy $,
   $W(|\alpha|^2)$ is a positive weight function
   may be found after specifying $f(n)$, $\hat I$ is the unity operator and $D$ is the domain of the states
   in the complex
   plane defined by the disk
  \begin{equation}\label{diskffe}
    D = \left\{\alpha \in \mathbb{C}, \; |\alpha| \leq \lim_{n\rightarrow \infty}
    [nf^2(n)]\right\},
  \end{equation}
    centered at the origin in the complex plane.
    Inserting the explicit form of the states (\ref{NCSs1}) in (\ref{res-id}) with $|\alpha|^2 \equiv x$
    it can be easily checked that the resolution of the identity
    holds if the following moment problem is satisfied:
  \begin{equation}\label{res-fffe}
      \pi \int _0^R dx \sigma (x) x^n=
      [nf^2(n)]!,\qquad n=0,1,2,\cdots ,
   \end{equation}
    where $\sigma(x) = \frac{W(x)} {N(x)}$ and $R$ is the radius of convergence determined by the relation
    (\ref{diskffe}). The condition (\ref{res-fffe}) presents a severe restriction on the
    choice of $f(n)$.
    In fact, only a relatively small number of  $f(n)$ nonlinearity
    functions are known for which the proper  functions $\sigma(x)$  exist.

    Comparing the dual pairs in (\ref{NCSs1}) and (\ref{dual}) it is readily found that requiring the resolution of
    the identity for the dual pair in (\ref{dual}), the following must
    hold
   \begin{equation}\label{res-ffff}
      \int_{\tilde D} d^2 \beta |\widetilde{\beta, f}\rangle \widetilde{W}(|\beta|^2)\langle\widetilde{\beta, f} | =
       \sum_{n=0}^\infty |n\rangle  \langle n| =\hat I,
   \end{equation}
   where $\widetilde{D}$ is the domain of the dual states
   in the complex plane defined by
  \begin{equation}\label{diskff}
   \widetilde{D} = \{\beta \in \mathbb{C}, \; |\beta|\leq \lim_{n\rightarrow \infty}
    [n/f^2(n)]\}.
  \end{equation}
  Setting $|\beta|^2 \equiv x$ and simplifying the left hand side of equation (\ref{res-ffff}) we are lead to the following moment integral
     \begin{equation}\label{res-fff}
      \pi \int _0^ {\widetilde{R}} dx \tilde \sigma (x) x^n=
      [n/f^2(n)]!,\qquad n=0,1,2,\cdots ,
   \end{equation}
    where $\widetilde{\sigma}(x) = \frac{\tilde W(x)} {\tilde N(x)}$ and $\widetilde{R}$ is the radius of convergence determined by the relation
    (\ref{diskff}).
We can introduce the "general solutions" of the moment problems
were introduced in (\ref{diskff}) and (\ref{res-fff}) which is
known as the inverse Mellin transform. Indeed  $\sigma (x) =
\mathcal{M}^{-1}[\rho_{\infty}(s-1); x]$ for the CSs defined in
the whole of space ($R=\infty$), and $ \sigma (x)H(R-x) =
\mathcal{M}^{-1}[\rho_R(s-1); x]$ for the states restricted to the
unit disk ($R<\infty$), where $H(z)$ is the Heaviside function and
we have defined $\rho(s-1)=[(s-1)f^2(s-1)]!$ (for a useful
treatment on Mellin and inverse Mellin transforms refer to
\cite{kps}
 and references therein). Similar expressions may be followed for $\tilde\sigma(x)$.



 \section{Constructing the combination of the dual family of nonlinear CSs}\label{sec-n2}

   Now we introduce the combination (first type of superposition) of nonlinear CSs with
   associated dual pair. In fact the explicit form of states can be easily obtained by
   linear combination of the un-normalized forms of (\ref{NCSs1}) and (\ref{dual})
   with the same nonlinearity function tends to the normalized state
 \begin{eqnarray}\label{superposed1}
  | \alpha,\beta,f\rangle &=& N_{s1} (\|\alpha,f\rangle+ \|\widetilde{\beta,f}\rangle)\nonumber \\
  &=& N_{s1}\sum_{n=\circ}^\infty\frac {{{\alpha}^n}+\beta^n\; ([f(n)]!)^{2}}{\sqrt{n!}\;[f(n)]!}\;| n  \rangle,
 \end{eqnarray}
  where the normalization factor $N_{s1}$ can be easily determined from the condition
  $\langle\alpha,\beta,f\vert\alpha,\beta,f\rangle=1$. Note that after dropping the
   normalization factor of $| \psi\rangle$ we have showed it by $\| \psi\rangle$, and
   the subscript  "$s1$" in (\ref{superposed1}) and what follows indicates the superposition of the first kind.
   Clearly taking $\beta=0$ or $\alpha=0$ in (\ref{superposed1}) leads to (\ref{NCSs1}) or (\ref{dual}),
   respectively. In this paper we restrict ourselves to the special case $\alpha=\beta$ and therefore
   the introduced state in (\ref{superposed1}) is simplified to
 \begin{equation}\label{superposed2}
   |\alpha,\alpha, f \rangle \equiv |\alpha,f_{s1}\rangle=N_{s1}\sum_{n=\circ}^\infty\frac {\alpha^n\;
    (1+([f(n)]!)^{2})}{\sqrt{n!}\;[f(n)]!}\;| n  \rangle,\qquad   \alpha \in \mathbb{C},
 \end{equation}
  where $N_{s1}$ may be determined as follows
 \begin{equation}\label{Ns}
   N_{s1}=\left[\sum_{n=\circ}^\infty\frac{|\alpha |^{2n}\;\left(1+([f(n)]!)^{2}\right)^{2}}
   {n!([f(n)]!)^{2}}\right]^{-\frac{1}{2}}.
 \end{equation}

    It is worth to mention that this form of superposition (called by us as combination) of a dual pair of nonlinear
    CSs can be classified in the following category of superpositions of nonlinear CSs as
    \begin{equation}\label{comb1}
 |\alpha, f_{s1} \rangle = c_{1} |\alpha, f\rangle+ c_{2} |\widetilde{\alpha, f}\rangle,
    \end{equation}
    where $c_{1}$ and $c_{2}$ are now real numbers determined as
    \begin{equation}\label{comb2}
      c_{1}=\frac{N_{s1}} {N_f}, \qquad c_{2}=\frac{N_{s1}}{\tilde{N}_f},
    \end{equation}
    depending on $f(n)$ and $\alpha$. Note that in this case $c_1$ and $c_2$ are
    constrained by the requirement of the normalizability of the combined states.
    The superposition proposed in (\ref{superposed2}) or equivalently (\ref{comb1})
    can be considered as the generalization of the superpositions of CSs to nonlinear
    CSs, referred  to in the introduction of the present paper
    \cite{prakash1, prakash2, prakash3, Wang, Nguyen}.
    We hope that these latter states also find their physical applications in relevant
    schemes on the performance of the teleportation protocols.

   It is well-known that there exists a simple relation between the expansion coefficients
   $C_{n}$'s of the nonlinear CSs with the corresponding nonlinearity function $f(n)$ as follows \cite{Manko1997}
  \begin{equation}\label{f(n)}
    f(n)=\frac{C_{n-1}}{\sqrt{n}C_{n}}.
  \end{equation}
  To verify the nonlinearity nature of $\vert\alpha,f_{s1}\rangle$ in (\ref{superposed2}),
  the relation (\ref{f(n)}) leads us to the following nonlinearity function expressed
  in terms of original nonlinearity function
  \begin{equation}\label{fs}
   f_{s1}(n)=\frac{1+\left([f(n-1)]!\right)^{2}}{1+\left([f(n)]!\right)^{2}}\;f(n).
  \end{equation}
  Therefore, by substitution $f(n)$ corresponding to any nonlinear oscillator algebra
  in (\ref{fs}) we are able to find $f_{s1}(n)$ associated to the special superposed
  state $\vert\alpha,f_{s1}\rangle$ introduced in (\ref{superposed2}).

   So such a superposition of any dual pair of states can in principle be obtainable
   as the eigenstate of the generalized annihilation operator $af_{s1}(n)=A_{s1}$
  \begin{equation}\label{A,Adag }
   A_{s1}|\alpha,f_{s1}\rangle=\alpha|\alpha,f_{s1}\rangle,\qquad  \alpha\in\mathbb{C},
  \end{equation}
  as follows
  \begin{equation}\label{superposed3}
   | \alpha,f_{s1}\rangle=N_{s1}\sum_{n=0}^{\infty}\frac{\alpha^n}{\sqrt{n!}[f_{s1}(n)]!}|n\rangle.
  \end{equation}
  In this way, a vast classes of nonlinear CSs may be obtained using various nonlinearity functions
  $f(n)$ in (\ref{superposed2}).
  It is also possible to consider the proposals of Roy {\it et al}
  \cite{roy roy} or Ali {\it et al} \cite{ ART} and reproduce the states in
  (\ref{superposed2}), (\ref{comb1}) or (\ref{superposed3}) via the action of a displacement-type operator
  $D_{s1}(\alpha) = \exp(\alpha C_{s1}^{\dag} - \alpha^{\ast} A_{s1})$ on the vacuum by
  $|\alpha,f_{s1}\rangle = D_{s1}(\alpha)|0\rangle,$
  where we have defined $C_{s1}=a\;\frac{1}{f_{s1}(n)}$
  which has the property $[ A_{s1}, C_{s1}^\dag]=\hat{I}$.
  It is a remarkable point that while previously superpositions of CSs
  for instance even and odd (linear \cite{dodonov} or nonlinear \cite{mancini}) CSs
  are not essentially remained in the $f$-deformed family of CSs, the
  proposed combined states in the present paper (as in the special superposition
  of nonlinear CSs recently introduced by us in \cite{superpos new}) are  exactly "$f$-deformed CSs".

  We end this section with mentioning that to the best of our knowledge in superposing the quantum
  states it is enough for one to have the main requirements of CSs  for the individual components of the superposed
  states \cite{dodonov, Schleish, mancini}
  (in this case the dual families).
  But, due to the existence of an explicit form
  of the nonlinearity function $f_{s1}(n)$ in (\ref{fs}) for the first kind of
  superpositions we hope that the resolution of the identity for the
  superposed states may be established in future works".

 \section{Superposition of a dual pair of nonlinear CSs}\label{sec-n3}
         \label{sec-n22}

 In this section we introduce the superposition (of the second type) of the two
 normalized components of the dual pair of any class of nonlinear CSs with known $f(n)$ and then normalize the
 resultant states. For this purpose, let us initially
 demonstrate the form of this superposition as
 \begin{equation}\label{sup1}
   |\psi\rangle _{s2}= N_{s2} (|\alpha,f\rangle+
   |\widetilde{\alpha,f}\rangle).
 \end{equation}
  We denoted the second type of superposition by $|\psi\rangle _{s2}$ to
  emphasis the fact that the resultant states are not $f$-deformed.
  Setting the relations (\ref{NCSs1}) and (\ref{dual}) in (\ref{sup1})
  we arrive at the explicit form of normalized superposition in the Fock space representation as
 \begin{equation}\label{sup2}
  |\psi\rangle _{s2} =N_{s2}\sum_{n=0}^\infty
  \frac {\alpha^n G(n, |\alpha|^2)}{\sqrt{n!}\;[f(n)]!}\;| n  \rangle,
 \end{equation}
  where for simplicity we have set  $G(n, |\alpha|^2) = N_{f}+\tilde{N}_{f}[f(n)!]^{2}$.
  By using $ _{s2}\langle \psi|\psi\rangle _{s2}=1$ one finds the normalization factor as
 \begin{equation}\label{sup3}
   N_{s2}= \left[\sum_{n=
   \circ}^\infty \frac {|\alpha|^{2n} G^2(n, |\alpha|^2)} {n!([f(n)]!)^2}\right]^{-\frac{1}{2}}.
 \end{equation}

  \section{non-classical criteria of states}\label{sec-n3}
   Generally, a state is known as a non-classical state (with no classical analogue) if
   the Glauber-Sudarshan $P(\alpha)$ function \cite{Sudarshan} can not be interpreted as a probability density. However,
   in practice one can not directly apply this criterion and this definition
   may hardly be applied to investigate the non-classicality nature of a
   state \cite {Shchulin-Richter_Vogel}. Altogether, this purpose has been mainly achieved
   customarily by checking the {\it "quadrature squeezing, amplitude squared squeezing,  sub-Poissonian
   statistics (antibunching) and oscillatory number distribution"}.
   A common feature of
   all the  above mentioned criteria is that  the corresponding
   $P$-function of a non-classical states is not positive definite.
   Therefore, each of the mentioned effects (first and second order squeezing or sub-Poissonian
   statistics which will be considered in the present paper) are sufficient
   condition for a state to exhibit non-classicality. It is worth mentioning that the necessary condition
   for the non-classicality of a states is yet the subject of recent researches \cite{Voegell}.

 \subsection{Sub-Poissonian statistics}

  To investigating the sub-Poissonian statistical behavior of the superpositions
  of states we will deal with second order
  correlation function \cite{Glauber2}
 \begin{equation}\label{g2(0)2}
   g^{2}(0)= \frac{\langle{a^\dag}^2\;a^2\rangle}{\langle{a^\dag}\;a\rangle^2}.
 \end{equation}
  Depending on the particular form of $f(n)$ the state may exhibit super-Poissonian,
  Poissonian or sub-Poissonian, respectively
  if $g^{2}(0)>1$ (bunching effect), $g^{2}(0)=1$ or $g^{2}(0)<1$ (antibunching effect).

   \subsection{Quadrature squeezing}

 In order to study quadrature squeezing we consider the following hermitian operators
  \begin{equation}\label{x y}
    x=\frac{a+a^\dag}{2},\qquad   y=\frac{a-a^\dag}{2i}.
  \end{equation}
 Then the uncertainty relation holds $\langle(\Delta x)^2\rangle
 \langle(\Delta y)^2\rangle\;\geq\frac{1}{16}$,
 where $\langle(\Delta x_{i})^2\rangle = \langle x_{i}^2 \rangle -
  {\langle x_{i} \rangle} ^2$, $x_{i}=x$ or $y$. A state is squeezed if any of the following conditions holds:
 \begin{equation}\label{delta xi2}
   \langle(\Delta x)^2\rangle  < \frac{1}{4} \qquad  or  \qquad \langle(\Delta y)^2\rangle <  \frac{1}{4}.
 \end{equation}
   Now by using equations (\ref{x y}) and (\ref{delta xi2}) the squeezing conditions
   may be transformed to the following inequalities:
 \begin{equation}\label{I1}
   I_{1}= \langle a^{2}\rangle+\langle{a^\dag}^{2}\rangle-{\langle a\rangle}^2-
   {\langle a^\dag\rangle}^2-2\;\langle a\rangle \;\langle a^\dag\rangle+2\;\langle a^\dag a\rangle < 0,
 \end{equation}
 and
 \begin{equation}\label{I2}
   I_{2}= -\langle a^{2}\rangle-\langle{a^\dag}^{2}\rangle+{\langle a\rangle}^2+
   {\langle a^\dag\rangle}^2-2\;\langle a\rangle \;\langle a^\dag\rangle+2\;\langle a^\dag a\rangle < 0,
  \end{equation}
    respectively for $x$ and $y$ quadratures.

  \subsection{Amplitude squared squeezing}

    In order to investigate the amplitude squared squeezing effect the following
    two hermitian operators have been introduced \cite{Hillery}
 \begin{equation}\label{X Y}
   X=\frac{a^2+{a^\dag}^2}{2},\qquad         Y=\frac{a^2-{a^\dag}^2}{2i}.
 \end{equation}
 In fact $X$ and $Y$ are the operators corresponding to the real and imaginary parts
 of the square of the complex amplitude of the electromagnetic field. Heisenberg
 uncertainty relation of these two conjugate operators is given by
 $\langle(\Delta X)^2\rangle\;\langle(\Delta Y)^2\rangle\;\geq \frac{1}{4}\;|\langle [X,Y] \rangle|^2.$
 It follows that our introduced states in (\ref{superposed2}) and (\ref{sup2}) will
 exhibit amplitude squared squeezing if
 \begin{equation}\label{deltaXi2}
   \langle(\Delta X)^2\rangle < \frac{1}{2}\;|\langle [X,Y] \rangle| \qquad
   \;or\;
   \qquad \langle(\Delta Y)^2\rangle < \frac{1}{2}\;|\langle [X,Y] \rangle|.
 \end{equation}
   With the help of equations (\ref{X Y}) and (\ref{deltaXi2}) one can introduce
   the following squeezing conditions corresponding to $X$ and $Y$, respectively
 \begin{eqnarray}\label{I3 I4}
    I_{3}&=& \frac{1}{4}\;\Big(\langle a^{4}\rangle+\langle{a^\dag}^{4}\rangle+\langle
     {a^\dag}^2 a^2\rangle+\langle a^2 {a^\dag}^2\rangle-
    {\langle a^2\rangle}^2-{\langle {a^\dag}^2\rangle}^2\nonumber\\
    &-& 2\;\langle a^2\rangle \;\langle {a^\dag}^2\rangle\Big)-\;\langle a^\dag a\rangle-\frac{1}{2} < 0,
 \end{eqnarray}
    and
 \begin{eqnarray}\label{I3 I4}
    I_{4}&=& \frac{1}{4}\;\Big(-\langle a^{4}\rangle-\langle{a^\dag}^{4}\rangle+\langle
    {a^\dag}^2 a^2\rangle+\langle a^2 {a^\dag}^2\rangle+{\langle a^2\rangle}^2+{\langle
    {a^\dag}^2\rangle}^2\nonumber\\
    &-& 2\;\langle a^2\rangle \;\langle {a^\dag}^2\rangle\Big)-\;\langle a^\dag a\rangle-\frac{1}{2}< 0.
  \end{eqnarray}
  It is evident that each of the expectation values in the formulas in this
  section must be evaluated with respect to a particular quantum state which is of
  interest (see appendix).


 \section{Physical realization}

  In this section we will apply the presented formalism to a generalized CSs with particular
  nonlinearity function. Then the previously mentioned non-classical signs  of the superposed
  states will be investigated graphically.
  For this purpose we consider the Hydrogen-like spectrum  which their shifted energy
  eigenvalues are given in  \cite{Hatom} by $ e_{n}=1-\frac{1}{(n+1)^2}$.
  The corresponding nonlinearity function of this system by considering the method proposed
  in \cite{Roknizadeh2004} may be written as
 \begin{equation}\label{f(n) H atom}
   f(n)= \frac{\sqrt{n+2}}{n+1},
 \end{equation}
   where the associated CSs have been defined in the unit disk centered at the origin.
   The resolution of the identity for the nonlinear CSs of Hydrogen-like spectrum and their dual pair
   has been recently established in \cite{GK} and \cite {Roknizadeh 2005},  respectively.
   Using (\ref{f(n) H atom}) in the mean values  of (\ref{g2(0)2})  (see Appendix)
   we have plotted second order
   correlation function for various states against $\alpha$ in figure 1.
   As it is clear, only the dual family of corresponding states according to (\ref{dual}) shows sub-Poissonian
   behavior, while for the introduced states $|\alpha,f_{s1}\rangle$ and $|\psi\rangle_{s2}$ as well as $|\alpha,f\rangle$,  $g^{2}(0)$
   is larger than unity (they exhibit super-Poissonian behavior). In fact comparing the curves displayed in figure 1 one may conclude that the quantum statistical behavior of $|\alpha,f_{s1}\rangle$
   and $|\psi\rangle_{s2}$ are qualitatively similar.
   From figures 2-a and 2-b it is found that quadrature squeezing
   occurred for $|\widetilde{\alpha,f}\rangle$ and $|\psi\rangle_{s2}$ in $x$, and for $|\alpha,f\rangle$
   as well as combined state $|\alpha,f_{s1}\rangle$ in $y$ quadrature.
   Clearly, to guarantee the Heisenberg uncertainty relation, the regions of squeezing occurrence in $x$ and $y$
   components for $|\psi\rangle_{s2}$ state is different from each other.
   Figures 3-a and 3-b show that amplitude squared squeezing has been
   occurred for $|\widetilde{\alpha,f}\rangle$ in $X$, and for
   $|\alpha,f\rangle$ as well as the combined state
   $|\alpha,f_{s1}\rangle$ in $Y$ component, while the superposed state $|\psi\rangle_{s2}$ does not show
   amplitude squared squeezing in $X$ nor in $Y$ component.
   Adding our numerical results for this special nonlinearity function, one may conclude
   that the behavior of the combination of states are qualitatively very similar to original state
   $|\alpha,f\rangle$, far from the dual states.

  \section{Superposition (of the first kind) of Gazeau-Klauder CSs with their dual pair}
    Adopting certain physical criteria rather than imposing selected mathematical requirements,
    Gazeau and Klauder by reparameterizing the generalized CSs $| \alpha\rangle$
    in terms of  two independent parameters $J$ and $\gamma$, introduced the generalized CSs
    $| J, \gamma\rangle$, known ordinarily as GKCSs in the physical literature \cite{GK}.
    On the other hand, in \cite{ed} the authors imposed a minor modification on these states via
    generalizing the Bargmann representation for the standard harmonic oscillator \cite{bargman}.
    In the present paper we proceed to consider this form of GKCSs for further superposition scheme
    of the first kind (combination).  The  {\it analytical representations} of GKCS associated to a
    physical system with the discrete nondegenerate spectrum $e_n$ have been introduced as follows
 \begin{equation}\label{gk}
  \vert z,\gamma\rangle ^{GK}=
  {N^{GK}}\sum_{n=0}^\infty \frac{z^n\;e^{-ie_{n}\gamma}}{\sqrt{\rho(n)}}\vert n\rangle,
  \qquad z \in \mathbb C, \qquad \;0\neq \gamma \in \mathbb R,
 \end{equation}
    where the normalization constant and the function $\rho(n)$ are given by
  \begin{equation}\label{N,p}
  {N^{GK}} = \left(\sum_{n=0}^{\infty} \frac{|z|^{2n}}{\rho(n)}\right)^{-\frac{1}{2}}, \qquad \rho(n)=[e_{n}]!.
  \end{equation}
    Also, the dual family of GKCSs as  the temporally stable CSs of the dual of KPS (Klauder-Penson-Sixdeniers) CSs \cite{kps},
    were introduced by one of us in \cite{Roknizadeh 2005}. Using the {\it analytical representation}
     the following form for the dual family has been deduced
 \begin{equation}\label{hgk}
    \vert \widetilde{z,\gamma}\rangle^{GK} = {{\tilde{N}}^{GK}}\sum_{n=\circ}^\infty
    \frac{z^n\;e^{-i\varepsilon_{n}\gamma}}{\sqrt{\mu(n)}}\vert n\rangle,
    \qquad z\in \mathbb C,\qquad 0\neq\gamma \in \mathbb R,
  \end{equation}
    where the normalization constant may be written as
  \begin{equation}\label{Nh,p}
     {{\tilde{N}}^{GK}} = \left(\sum_{n=0}^{\infty} \frac{|z|^{2n}}{\mu(n)}\right)^{-\frac{1}{2}},
  \end{equation}
    and $\mu(n)$ and $\varepsilon_{n}$ are  respectively given by
 \begin{equation}\label{mu..}
   \mu(n) = \frac{(n!)^2}{\rho(n)},\qquad    \varepsilon_{n} = \frac{n^2}{e_{n}}.
 \end{equation}
   In (\ref{mu..}) $ \varepsilon_{n}$ denotes the energy spectrum of the Hamiltonian associated to the dual pair states.
   The dual family which will constitute the two components of our
   superposition in the following subsection are not orthogonal, with
   the overlap
 \begin{equation}\label{overlap2}
  {}^{GK}\langle z,\gamma | \widetilde {z,\gamma}\rangle ^{GK} =
  {N^{GK}}{{\tilde{N}}^{GK}}\sum_{n=0}^{\infty}\frac{|z|^{2n}e^{i(e_{n}-\varepsilon_{n})\gamma}}{n!}.
 \end{equation}
 A few words seems to be useful about the over-completeness of
 these states. In this relation it is a clear fact that the states belong to each pair of the dual family are indeed nonorthogonal.
 Also, the over-completeness of GKCSs requires satisfying the resolution of the identity according to:
   \begin{equation}\label{res-gkcs}
      \int _0 ^R d \mu(z) |z, \gamma \rangle ^{GK} \; {}^{GK}\langle z, \gamma| = \sum_{n=0}^\infty |n\rangle  \langle n|
      =\hat I.
   \end{equation}
    Inserting the explicit form of the states (\ref{gk}) in (\ref{res-gkcs}) with $|z|^2 \equiv x$
    simplifies it to  the following moment problem:
  \begin{equation}\label{res-fgk}
      \pi \int _0^R dx \sigma (x) x^n=
      \rho(n),\qquad n=0,1,2,\cdots.
   \end{equation}
   where $R$ is the radius of convergence may be determined similar to nonlinear CSs.
   The condition (\ref{res-fgk}) again presents a severe restriction on the
   choice of $\rho(n)$ and consequently $e_n$.
   Following the same path one may immediately obtain the conditions for the
   over-completeness of the dual pair of GKCSs.


 \subsection{Mathematical structure of the combination of the dual pair of GKCSs}

    Now by using the un-normalized form of equations (\ref{gk}) and (\ref{hgk}),
    we construct superposition of a dual pair of GKCSs as
    $\vert z,\gamma\rangle_{s1}^{GK}= N_{s}^{GK}(\| z,\gamma\rangle^{GK} + \| \widetilde{z,\gamma}\rangle^{GK}).$
    It is straightforward to obtain its explicit representation in Fock space as
 %
  \begin{equation}\label{gks2}
   {\vert z,\gamma\rangle}_{s1}^{GK}= N_{s1}^{GK}\sum_{n=0}^\infty \frac{z^n}{n!}
   \frac{K(n, \gamma)}{\sqrt{\rho(n)}}|n\rangle,
 \end{equation}
  where we have introduced $K(n, \gamma)= n!\; e^{-i\gamma e_{n}}+[e_{n}]!\;e^{-i\gamma \varepsilon_{n}}$.
  The normalization constant $N_{s1}^{GK}$ can be obtained as follows
 \begin{equation}\label{NsGK}
   N_{s1}^{GK}=\left(\sum_{n=0}^{\infty}\frac{|z|^{2n}|K(n, \gamma)|^{2}}{[e_{n}]!(n!)^2}\right)^{-\frac 1 2}.
 \end{equation}
 The nonlinearity of the combined states in (\ref{gks2}) may be established with the result
  \begin{equation}\label{fgk}
    f_{s1}^{GK}(n, \gamma)= \frac{K(n-1, \gamma)}{K(n, \gamma)},
  \end{equation}
   and by convention $ [f_{s1}^{GK}(0, \gamma)]!\doteq 1$. Note that on the contrary to the
   previous combination of the dual pair of nonlinear CSs which perfectly remained
   in the $f$-deformed states,  $\vert z,\gamma\rangle_{s1}^{GK}$ in (\ref{gks2}) are not
   of GKCSs type, i.e., they are not temporally stable.

   As one may recognize the superposition of the second kind of GKCSs need more calculations,
   but yet with the same procedures of section 5, so we ignore studying it at  present.

 \subsection{Physical realization}

   Now we use the results of previous subsection to study the non-classical
   properties of P\"{o}schl-Teller potentials \cite{push}. It must be mentioned that
   this physical system is widely used in quantum mechanics specially in atomic and molecular physics.
   The energy spectrum of such system is given by $e_{n}= n(n+\lambda+\kappa)$,
   where $\lambda,\kappa >1$ characterize the potential. Following the path of \cite{Roknizadeh2004}
   the corresponding nonlinearity function is obtained as
  \begin{equation}\label{push en}
    f(n)= \sqrt{n+\lambda+\kappa}.
  \end{equation}
   The resolution of the identity for the GKCS and their dual pair associated to P\"{o}shl-Teller
   potential has been recently  established in \cite{push} and \cite{Roknizadeh 2005}, respectively.
   Now by substitution (\ref{push en}) in expectation values needed for the non-classical behavior (see Appendix),
   we can investigate the non-classical properties of a dual pair of GKCSs ((\ref{gk}) and (\ref{hgk})) and the
   combination of states introduced in (\ref{gks2}).
   In figure 4 we have plotted the curve of second order correlation function in (\ref{g2(0)2}) against $z$ for
   fixed values $\gamma=0.5$ and $\lambda= 4 =\kappa$. As an interesting feature of our numerical results it is clear
   that the interference effects in the combination of the dual pair of GKCSs vanish the sub-Poissonian behavior
   of the individual GKCSs and its dual pair in all ranges of $z$, i.e., the state in
   (\ref{gks2}) shows super-Poissonian statistics.
  Figures 5 and 6 show quadrature squeezing and amplitude
  squared squeezing for the three sets of states, original GKCSs $|z, \gamma \rangle^{GK}$, their dual pair
  $|\widetilde{z,\gamma}\rangle^{GK}$ and the combination of states
  $|z,\gamma\rangle_{s1}^{GK}$, respectively. In all cases we have choosed $z\in \mathbb{R}$.
  From figures 5-a, 5-c and 5-e for squeezing in $x$ quadrature one can see that the original GKCSs,
  shows oscillatory behavior, no squeezing for the dual pair as well as the combined state.
  On the other side from figures 5-b, 5-d and 5-f the oscillatory behavior of squeezing in $y$
  quadrature is seen for the original GKCSs, squeezing in all ranges of $\gamma$
  and $z$ for the dual pair as well as the combined state.
  Interestingly, similar qualitatively behavior in $X$ may be observed from figures 6-a, 6-c
  and 6-e, and in $Y$ from figures 6-b, 6-d and 6-f when amplitude
  squared squeezing is evaluated.

 \section{Summary and Conclusion}

 In summary we introduced the superposition of the dual pair of $f$-deformed states
 in two distinct approaches.
 The presented formalisms have been applied to a familiar physical system, i.e.,
 Hydrogen-like spectrum. We continued the work with the dual pair of GKCSs, the states
 which were extracted from a more physical insights. For the latter states we have used the
 P\"{o}schl-Teller potentials as a physical realization.
 Adding our results some considerable points are remarkable. Firstly, the combination
 (superposition of the first type) of any dual pair of nonlinear CSs and GKCSs maintained
 them in the family of nonlinear CSs, although with a modified $f_{s1}(n)$ and $f_{s1}^{GK}(n, \gamma)$
 rather than original $f(n)$ and $f^{GK}(n, \gamma)$, respectively. This observation is unlike the usual
 types of superpositions more frequently have been introduced in the literature, also has been
 done by us in section 4 of the present paper (for instance even and odd, for standard and nonlinear CSs).
 Finding the explicit forms of $f_{s1}(n)$ and $f_{s1}^{GK}(n, \gamma)$ has the potentiality of the
 reconstruction the combination of states via the two well-known  approaches: i)
 the algebraic method and ii) group theoretical procedures.
 Secondly, it is noticeable that  by considering the formalism introduced by one us in \cite{Roknizadeh2004},
 we can give Hamiltonians describe the dynamics of the systems corresponding to the combined states $|\alpha, f_{s1}\rangle$ in (\ref{fs}) and $\vert z,\gamma\rangle_{s1}^{GK}$  in (\ref{fgk}). Indeed, using (\ref{fs}) the $f$-deformed ladder operators formalism lead one to the Hamiltonian $\hat{H}_{s1}=A^{\dag}_{s1}A_{s1}=nf^{2}_{s1}(n)$, and using (\ref{fgk}) to $\hat{H}_{s1}^{GK}={A^{GK}_{s1}}^{\dag} A_{s1}^{GK} = n |f^{GK}_{s1}(n, \gamma)|^2$.

 Thirdly, the superpositions in the present paper do not always highlight the non-classical
 effects, there is special cases in which it weaken the non-classical properties.
 This is not surprising if one recalls that the number states are with highest non-classicality
 and canonical CSs (which are particular superposition of the number states) are with lowest
 non-classicality  (and highest classicality).  To this end we are willing to mention that the
 presented formalism can be easily extended to any physical system, either with known nonlinearity
 function or with known discrete non-degenerate spectrum
 (center of mass motion of trapped ion, harmonious states,
 anharmonic oscillator and generally  any nonlinear oscillator algebra).


 \section{Appendix}

 Considering the introduced states in (\ref{superposed2}), the expectation values needed
 for the calculation and computation of the non-classicality criteria have been expressed
 in section 5 corresponding to the first kind of superposition (combination) of states can be easily obtained as

 \begin{equation}\label{a}
     \langle a\rangle_{s1}=|N_{s1}|^2\sum_{n=0}^\infty
     \frac {\alpha^{n+1}\;{\alpha^\ast}^{n}\;\left(1+([f(n+1)]!)^2)
     \right)\left(1+([f(n)]!)^2)\right)}{{n!}\;\;[f(n)]!\;\;[f(n+1)]!},
 \end{equation}

 \begin{equation}\label{a2}
     \langle a^2\rangle_{s1}=|N_{s1}|^2\sum_{n=0}^\infty
     \frac {\alpha^{n+2}\;{\alpha^\ast}^{n}\;\left(1+([f(n+2)]!)^2)\right)
     \left(1+([f(n)]!)^2)\right)}{{n!}\;\;[f(n)]!\;\;[f(n+2)]!},
    \end{equation}

 \begin{equation}\label{a4}
     \langle a^4\rangle_{s1}=|N_{s1}|^2\sum_{n=0}^\infty
     \frac {\alpha^{n+4}\;{\alpha^\ast}^{n}\;\left(1+([f(n+4)]!)^2)\right)\;\left(1+([f(n)]!)^2)
     \right)}{{n!}\;\;[f(n)]!\;\;[f(n+4)]!},
 \end{equation}

  \begin{equation}\label{n}
  \langle a^\dag a\rangle_{s1}= |N_{s1}|^2\sum_{n=0}^\infty
   \frac {|\alpha|^{2(n+1)}\;\left(1+([f(n+1)]!)^2)\right)^{2}}{{n!}\;\;([f(n+1)]!)^2},
 \end{equation}

   \begin{equation}\label{ad2a2}
  \langle {a^\dag}^2 a^2\rangle_{s1}= |N_{s1}|^2\sum_{n=0}^\infty
   \frac {|\alpha|^{2(n+2)}\;\left(1+([f(n+2)]!)^2)\right)^{2}}{{n!}\;\;([f(n+2)]!)^2},
  \end{equation}


   Similarly, the expectation values needed for the calculation and computation of
   the non-classicality criteria corresponding to the second kind of superposition of
   states introduced in (\ref{sup2}) are as follows

    \begin{equation}\label{sup7}
     \langle a\rangle_{s2}=|N_{s2}|^2\sum_{n=0}^\infty
    \frac {\alpha^{n+1} {\alpha^{\ast}}^{n}\;G(n+1, |\alpha|^2) G(n, |\alpha|^2)}{{n!}\;\;[f(n)]![f(n+1)]!},
     \end{equation}

    \begin{equation}\label{sup6}
     \langle a^2\rangle_{s2}=|N_{s2}|^2\sum_{n=0}^\infty
    \frac {\alpha^{n+2} {\alpha^{\ast}}^{n}\;G(n+2, |\alpha|^2) G(n, |\alpha|^2)}{{n!}\;\;[f(n)]![f(n+2)]!},
     \end{equation}

     \begin{equation}\label{sup8}
    \langle a^4\rangle_{s2}=|N_{s2}|^2\sum_{n=0}^\infty
    \frac {\alpha^{n+4} {\alpha^{\ast}}^{n}\;G(n+4, |\alpha|^2) G(n, |\alpha|^2)}{{n!}\;\;[f(n)]![f(n+4)]!},
     \end{equation}

    \begin{equation}\label{sup5}
   \langle a^\dag a \rangle_{s2}=|N_{s2}|^2\sum_{n=0}^\infty
    \frac {|\alpha|^{2(n+1)}\;G^2(n+1, |\alpha|^2)}{{n!}\;\;([f(n+1)]!)^2},
    \end{equation}

    \begin{equation}\label{sup4}
   \langle {a^\dag}^2 a^2 \rangle_{s2}=|N_{s2}|^2\sum_{n=0}^\infty
    \frac {|\alpha|^{2(n+2)}\;G^2(n+2, |\alpha|^2)}{{n!}\;\;([f(n+2)]!)^2},
    \end{equation}


   To investigate the non-classical properties of GKCSs introduced in (\ref{gks2}), we need the following mean values
   \begin{equation}\label{agk}
    \langle a\rangle _{s1}^{GK}=\vert{N_{s1}}^{GK} \vert ^2\;\sum_{n=0}^\infty \frac
    {z^{n+1}{z^{\ast}}^{n} K({n+1}, \gamma) {K^{\ast}(n, \gamma)}}{(n!)^2 \sqrt{[e_{n}]!\;[e_{n+1}]!\;(n+1)}},
   \end{equation}

   \begin{equation}\label{a2gk}
     \langle a^{2}\rangle  _{s1}^{GK} = \vert{N_{s1}}^{GK} \vert ^2\;\sum_{n=\circ}^\infty
      \frac{z^{n+2}{z^{\ast}}^{n} K({n+2, \gamma}) {K^{\ast}(n, \gamma)}}{(n!)^2 \sqrt{[e_{n}]!\;[e_{n+2}]!\;(n+1)(n+2)}},
  \end{equation}

  \begin{equation}\label{a4gk}
   \langle a^{4}\rangle  _{s1}^{GK}=\vert{N_{s1}}^{GK} \vert ^2\sum_{n=\circ}^\infty
   \frac{z^{n+4}{z^{\ast}}^{n} K({n+4}, \gamma) {K^{\ast}(n, \gamma)}}{(n!)^2 \sqrt{[e_{n}]!\;[e_{n+4}]!\;(n+1)\cdots(n+4)}},
  \end{equation}

  \begin{equation}\label{adagk}
    \langle a^\dag a\rangle  _{s1}^{GK} =\vert{N_{s1}}^{GK} \vert ^2\;\sum_{n=\circ}^\infty
    \frac{{\vert z\vert ^{2n}} \vert K(n, \gamma)\vert ^{2}}{(n!)^2\;[e_{n}]!}\;n,
 \end{equation}

 \begin{equation}\label{ad2a2gk}
   \langle {a^\dag}^{2}a^2\rangle _{s1}^{GK} =\vert{N_{s1}}^{GK} \vert ^2\;\sum_{n=\circ}^
   \infty \frac{{\vert z\vert ^{2n}} \vert K(n, \gamma)\vert ^{2}}{(n!)^2\;[e_{n}]!}\;n(n-1).
 \end{equation}

   It is clear that the mean values of $\langle a^\dag\rangle_{si}$,
   $\langle{a^\dag}^{2}\rangle_{si}$,
   $\langle{a^\dag}^{4}\rangle_{si}$  and $\langle a^{2}{a^\dag}^{2}\rangle_{si}$
   can be obtained by taking  the
   conjugates of $ \langle a\rangle_{si}$, $\langle a^{2}\rangle_{si}$, $\langle a^{4}\rangle_{si}$ and
   $\langle {a^\dag}^{2} a^{2}\rangle_{si}$, respectively.
   Note that $i$ in  the subscripts $si$  may be $1$ or $2$ respectively for the first and second type
   of superpositions.

  {\bf Acknowledgement:} We would like to thank the referees for
       their helps in improving the quality of the paper.

 \include{thebibliography}


 \vspace {3 cm}
 \newpage
 {\bf FIGURE CAPTIONS}

 \vspace{1cm}

 {\bf FIG. 1}
       The curves of second order correlation function ${g^2}(0)$ corresponding to
       the states $| \alpha,f \rangle$, $| \widetilde{\alpha, f}\rangle $,
       $| \alpha,f_{s1} \rangle$ and $| \psi  \rangle _{s2}$ against amplitude
       $\alpha  \in \mathbb{R}$ for Hydrogen-like spectrum.

 \vspace {.5 cm}

 {\bf FIGs. 2}
       The plots of the inequality $I_{1}$ indicates squeezing in $x$ quadrature
       according to equation (29) displayed in 2-a, and the inequality $I_{2}$  indicates squeezing
       in $y$ quadrature according to equation (30) displayed in 2-b corresponding to
       the states $| \alpha,f \rangle$, $| \widetilde{\alpha, f}\rangle $,
       $| \alpha,f_{s1} \rangle$ and $| \psi  \rangle _{s2}$  against amplitude $\alpha \in \mathbb{R}$
       for Hydrogen-like spectrum.

 \vspace {.5 cm}

 {\bf FIGs. 3}
     The plots of the inequality $I_{3}$ indicates amplitude squared squeezing
     in $X$ according to equation (33) displayed in 3-a, and the inequality $I_{4}$
     indicates amplitude squared squeezing in $Y$ according to equation (34)
     displayed in 3-b corresponding to the states $| \alpha,f \rangle$, $| \widetilde{\alpha, f}\rangle $ ,
     $| \alpha,f_{s1} \rangle$ and $| \psi  \rangle _{s2}$ against $\alpha \in \mathbb{R}$ for
     Hydrogen-like spectrum.

 \vspace {.5 cm}

  {\bf FIG. 4}
                 The curves of second order correlation function ${g^2}(0)$  corresponding to the
                 states $| z, \gamma \rangle^{GK}$, $| \widetilde{z, \gamma}\rangle^{GK}$ and
                 $| z, \gamma \rangle_{s1}^{GK}$  against $z \in \mathbb{R}$
                 for P\"{o}schl-Teller potential at fixed $\gamma=0.5$ and $\lambda=4=\kappa$.

 \vspace {.5 cm}

  {\bf FIGs. 5}
                 The three-dimensional plots of the inequality $I_{1}$ for squeezing in $x$ quadrature according to
                 equation (29) against $z$ and $\gamma$ using the
                 states $| z, \gamma \rangle^{GK}$, $| \widetilde{z, \gamma}\rangle^{GK}$ and
                 $| z, \gamma \rangle_{s1}^{GK}$, displayed respectively in  figures 5-a, 5-c and 5-e,
                 and the inequality $I_{2}$
                 for squeezing in $y$ quadrature according to equation (30) against $z$ and $\gamma$ for the
                 above three sets of states in figures 5-b, 5-d and 5-f, respectively. In all cases the
                 P\"{o}schl-Teller potential is considered for $\lambda=4=\kappa$.
 \vspace {.5 cm}

  {\bf FIGs. 6}
                  The three-dimensional plots of the inequality $I_{3}$ for amplitude squared squeezing in $X$
                  according to equation (33) against $z$ and $\gamma$ using the
                  states $| z, \gamma \rangle^{GK}$, $| \widetilde{z, \gamma}\rangle^{GK}$ and
                  $| z, \gamma \rangle_{s1}^{GK}$, displayed respectively  in  figures 6-a, 6-c and 6-e,
                  and the inequality $I_{4}$ for amplitude squared squeezing in $Y$
                  according to equation (34) against $z$ and $\gamma$ for the above three sets of states
                  in figures 6-b, 6-d and 6-f, respectively. In all cases the P\"{o}schl-Teller potential
                  is considered with $\lambda=4=\kappa$.
 \vspace {.5 cm}

\end{document}